\documentstyle[prl,aps]{revtex}

\begin{document}

\input{epsf}

\twocolumn[\hsize\textwidth\columnwidth\hsize\csname@twocolumnfalse\endcsname

\title{Collective oscillations driven by correlation
in the nonlinear optical regime}

\draft

\author{Th. \"Ostreich\cite{goe} and L. J. Sham\cite{lajolla}}

\address{Department of Physics, University of California Berkeley and
Lawrence Berkeley National Laboratory,
Berkeley, CA 94720}

\date{\today}

\maketitle

\begin{abstract}
We present an analytical and numerical study of
the coherent exciton polarization including
exciton-exciton correlation. The time evolution
after excitation with ultrashort optical pulses
can be divided into a slowly varying polarization
component and novel ultrafast collective modes.
The frequency and damping of the collective modes are
determined by the high-frequency properties of
the retarded two-exciton correlation function,
which includes Coulomb effects beyond the mean-field approximation.
The overall time evolution depends on the low-frequency
spectral behavior.
The collective mode, well separated from the slower coherent density
evolution, manifests itself in the coherent emission of a
resonantly excited excitonic system, as demonstrated numerically.
\end{abstract}

\pacs{PACS numbers: 71.35.Lk, 42.50.Md, 78.47.+p}

\vskip2pc]
\vskip 0.1 truein

\narrowtext

% ----------------------------------------------------------------

Observations of coherent optical phenomena require a sufficiently
long dephasing time.
Thus, coherent processes such as the Rabi oscillations \cite{rabi} and
form-invariant pulse propagation \cite{hahn}  (also
known as self-induced transparency) are well studied both
theoretically and experimentally
in atomic vapors \cite{allen}.
In solid-state physics, the
nonlinear optical properties of semiconductors
are expected to exhibit analogous behavior, especially under
resonant excitation of excitons \cite{shah,chemla}, where
the exciton transition approximates an ideal two-level system.
However, there is a distinct difference in semiconductors. At moderate
excitation densities, the mobile, large-radius Wannier excitons suffer strong
Coulomb interaction on close approach and, at  high densities, the
collection of excitons dissociate into a neutral plasma.

A mean-field theory of the exciton-exciton interaction
leads to a purely {\it coherent} repulsive interaction, which
does not yield a mechanism for dephasing due to the fluctuating effective
local field. This level of approximation allows for  nonlinear electric
field-driven density oscillations,  which are termed as Rabi oscillations
\cite{binder,exact,schul}.  A further low-density approximate relation for the
density and the polarization leads to a Gross-Pitaevskii equation
\cite{gross,pita} for the exciton dynamics.  This yields form-invariant
soliton solutions \cite{hana,talanina,giess}.

Keldysh and Kozlov \cite{KK} included correlation in a
low-density ensemble of excitons. Here we
present an extension of the Gross-Pitaevskii equation
to include the correlation effects and explore some
fundamental consequences in the ultrafast nonlinear optical processes.
The correlation effects are based on our previous theory  of exciton
correlation in the weak nonlinear optical regime and are contained entirely
in a {\it retarded memory function} \cite{review,avk}.
The reactance part of the correlation function provides a renormalization of
the  coherent interaction and the admittance part provides a dephasing
mechanism from exciton-exciton interaction. The latter contribution may
be regarded as a nonlocal temporal excitation-induced dephasing \cite{eid}.
  A new finding is the ultrafast collective density oscillations riding on the
slow coherent density, both being excited by a short laser pulse.  That this
oscillation survives the excitation pulse shows that it is not related to the
Rabi oscillations.  Because of the inclusion of the correlation effects,
which, related by a sum rule to the exciton local-field interaction, must be
just as strong, the collective oscillations and the slow coherent dynamics
together are expected to give a more accurate picture of interacting excitons
in the density regime below dissociation.

% ----------------------------------------------------------------

The coherent dynamics of the exciton \cite{review} can be described
by a nonlinear equation of motion for the
polarization density $P$ of the form:

\begin{eqnarray}  \label{main}
  i \dot{P} &=&   \left( \delta  + \frac{\beta}{2} |P|^2 \right) P
 - \frac{1}{2}
 \left( 1- \frac{ |P|^2}{n_c} \right) \Omega
\\ &-&  \frac{i P^\ast}{2} \int_{-\infty}^t F(t-t') P(t')^2 dt' \, .
\nonumber
\end{eqnarray}

The equation is easily generalized to include
the spin, higher exciton levels, and the continuum and to account for spatial
variation. It is then exact to the third order in the exciting field
\cite{review,axt,diag1}.  The whole equation is thus valid for the low-density
regime.  Our purpose here is to explore the temporal variation of the exciton
dynamics at low excitation density and, therefore, we retain only the
resonantly  excited exciton transition and set
$P$ to be the usual polarization density divided by
$\mu |\phi_{1s}(0)|^2$, $\mu$ being the dipole moment and $\phi_{1s}(r)$ the
relative motion wave function.    For the terms in order on
the right side of the equation,
$\delta$ denotes the detuning of the central laser  frequency from the
resonance, $\beta>0$ is a measure of the mean-field exciton-exciton
interaction,  and $n_c$ the maximum density divided by
$|\phi_{1s}(0)|^2$ for the validity limit in the
Pauli blocking term.  The Rabi frequency
$\Omega$ is given by the dipole energy of the laser field amplitude.
Exciton-exciton correlation is contained in the retarded
memory function $F(\tau)$.  Similar effective equations without the
correlation term $F$ were
already used in the literature \cite{chemla,exho,non}.  A simple way to
account for the correlation effect by connecting the polarization
equation to an antiparallel-spin biexciton density \cite{kner} is
equivalent to setting $F$ to be the single-frequency biexciton propagator.

The retarded memory function is given by \cite{review},
\begin{eqnarray} \label{memory}
  F(\tau) = V |\phi_{1s}(0)|^2 \langle 0| D e^{-iH \tau}
  D^\dagger  |0 \rangle \; \Theta(\tau)\;,
\end{eqnarray}
in terms of the Hamiltonian of the system $H$, volume $V$, and the interaction
operator between two excitons,
\begin{eqnarray}  \label{dd}
  D &=& \int dr_1\int dr_2\int dr'_1\int dr'_2 B(r_1,r_2)
[v(r_1 - r'_1) \nonumber \\
 &+& v(r_2 - r'_2) - v(r_1 - r'_2) -  v(r_2 -
r'_1)]B(r'_1,r'_2),
\end{eqnarray}
with the Coulomb interaction $v(r)$ and the exciton annihilator
\begin{eqnarray} \label{aa}
  B(r_1,r_2) = \phi_{1s}(r_1 - r_2) \psi^{\dagger}_1(r_1)\psi_2(r_2),
\end{eqnarray}
where $\psi^{\dagger}_1, \psi_2$ are, respectively, the electron creator in
the  valence band and the annihilator in the  conduction band.
The formulas in configuration space show clearly the inclusion of the
electron-hole attraction in each exciton, leaving only the interaction between
the constituents of the two excitons in the interaction operator $D$.  The
correlation effects contained in the memory function are in second or higher
order of the interaction between two excitons.

Our numerical simulation below shows that intrinsic dephasing times due
to exciton-exciton scatterings are of the order of a few picoseconds in
a GaAs quantum well, of the same order as that due to electron-phonon
\cite{chemla}.  Thus, the conclusions drawn from the exclusion of  the
extrinsic dephasing would not be qualitatively changed by its inclusion.
The equation for the  polarization density resembles the
Gross-Pitaevskii equation for a single component Bose-Einstein  condensate
\cite{gross,pita}.  However, there are two essential differences, (1)
the exciton coherence is driven externally, and (2)
the effective interaction between polarization includes the
correlation term beyond the mean-field theory, which may well be applied to
the trapped atoms as their density increases.

After excitation with a short laser pulse at time $t=0$, the coherent
polarization evolves without the driving field $\Omega$. We present first an
analytic solution of Eq.~(\ref{main}) under the  assumption of a sufficient
slowly varying polarization $P_0(t)$,  which is then used to understand the
behavior of the numerical solutions without this approximation. Under the
slowly varying assumption, the rapidly decaying memory function can be
decoupled from the polarization
\begin{eqnarray}  \label{markoff}
  \int_{-\infty}^t F(t-t') P_0^2(t')^2 dt' \approx
  P_0^2(t) \int_{0}^{t} F(t-\tau) d\tau  \, .
\end{eqnarray}
This is similar to a Markov approximation for the polarization but the
scattering rate is still a time-dependent memory function, which we group
with the mean-field term from Eq.~(\ref{main})  in a new function:
\begin{eqnarray}  \label{gfunc}
  g(t) = \int_{0}^{t} F(\tau) d\tau + i \beta \, .
\end{eqnarray}
After the exciting pulse dies down, the detuning $\delta$ term can be included
in $P_0$ and removed from the equation of motion. For the resultant equation,
the ansatz $P_0 = n^{1/2}e^{i\phi}$ solves the slow time dynamics:
\begin{eqnarray}  \label{solution}
  n(t) &=& n_0 \left(1 + n_0 \int_0^t g_R(t') dt' \right)^{-1} ,
\nonumber \\
  \phi(t) &=& \phi_0 - \frac{1}{2} \int_0^t  g_I(t') n(t') dt' ,
\end{eqnarray}
for the initial conditions $n(0)=n_0, \phi(0)=\phi_0$ set by the exciting
pulse.
The real part of $g(t)$,  $g_R(t)$, determines the dephasing of the coherent
polarization and, thus, the exciton population.  The imaginary part, $g_I(t)$
determines the phase of the polarization. The mean-field effects are included
 in the total memory function $g(t)$.

The normalized spectral density of the $D$-$D$ correlation function, denoted
by $\rho(\omega)$, is $1/\pi$ times the real part of the Fourier transform of
$F(\tau)/F(0)$.  The sum rule \cite{rules},
\begin{eqnarray}  \label{sumrule}
  F(0) \int_0^\infty \omega^{-1} \rho(\omega)  d\omega = \beta
\end{eqnarray}
shows important cancellation in $g(t)$ between the mean-field exchange and the
correlation effects.
The low-frequency behavior of the spectral density is
governed by a power law,
$\rho(\omega) \sim \omega^\alpha$, discussed below.  It determines the
asymptotic long-time dynamics via:
\begin{eqnarray}  \label{longtime}
  g(t) \sim i \int_{0}^\infty \omega^{\alpha-1}
  e^{-i \omega t} d\omega
  = i^{\alpha+1} \Gamma(\alpha) t^{-\alpha},
\end{eqnarray}

\begin{figure}[h]
\epsfxsize7.0cm
\epsfbox{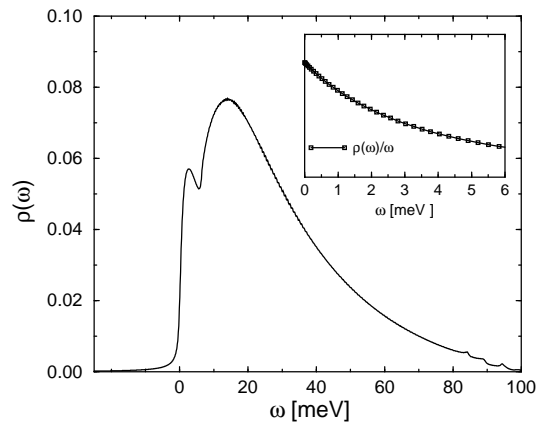}
  \caption{Equal-spin correlation function
    (solid-line) for a one-dimensional model system.
    The inset shows the exact spectral density $\rho(\omega)/\omega$
    for small $\omega$, indicating a linear dependence on $\omega$.
     \label{f1}}
\end{figure}
where $\Gamma$ is the usual Gamma function,
leading to a crossover
from a decaying coherence for $\alpha \le 1 $ to a constant density
$n(\infty) \le n_0$ for $\alpha > 1$.
In Figure~\ref{f1}, the spectral density of the equal-spin
correlation function is plotted for a one-dimensional system defined in
\onlinecite{review} with the exciton binding energy $\omega_x = $ 6.7~meV, the
one-electron bandwidth of       50~meV and equal electron and hole mass. An
artificial  broadening of 0.4~meV is used to smooth the energy splittings due
to the finite simulation size of $N=320$ sites. The inset shows the exact
behavior of the spectral density for small $\omega$, indicating a linear
behavior, i.e., $\alpha=1$.

\begin{figure}[h]
\epsfxsize7.5cm
\epsfbox{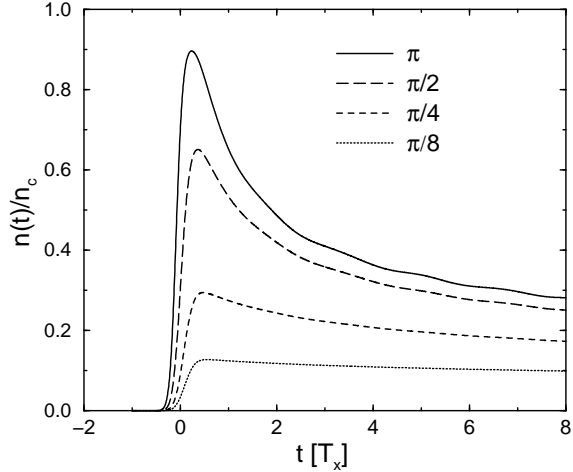}
  \caption{
    Coherent exciton density for the one-dimensional model
of Fig.~\protect{\ref{f1}} for a pulse intensity FWHM of 200~fs
and increasing pulse area $\Theta=\int \Omega(\tau) d\tau$.
The time unit $T_x= 2\pi/\omega_x$. \label{f2}}
\end{figure}

Figure~\ref{f2} shows the result of the numerical solution of Eq.~(\ref{main})
for the coherent evolution of the exciton density for the model in
Figure~\ref{f1}, using the derived values of $\beta=$ 6.6~meV, $F(0)=$ 48~meV
and $n_c=0.38$. The transient time evolution under the excitation with a
200~fs pulse (intensity FWHM) of increasing  area shows a hint of density
oscillations.  The pronounced decay of the population shortly after the
excitation is due to the decoherence of the exciton-exciton interaction,
following by a weak decay asymptotic behavior of the $\alpha=1$ case discussed
above. This regime will also be affected by dephasing of the exciton
due to additional degrees of freedom not included in the model.
Accurate correlation functions for realistic quantum-wells or bulk systems are
not yet available. We construct a model correlation
function for two equal-spin excitons in a two-parabolic-band model for a
three-dimensional semiconductor.
In systems of higher dimensions than one, we expect, by a phase space argument
for the scattering of two excitons, that the low-frequency dependence of the
spectral density is super-linear.  Combining a low-frequency power law with
a high-frequency decay  approximated as exponential, we  arrive at a simple
model for the spectral density
\begin{eqnarray}  \label{model}
  \rho(\omega) =
  \frac{\omega^\alpha exp(-\omega/\omega_F) }{\Gamma(\alpha+1)
    \omega_F^{\alpha+1}} ,
\end{eqnarray}
for positive frequencies with a single parameter $\alpha(>0)$. The values for
$\beta=52\omega_x/3$, $1/n_c= 7$, and $F(0)\approx
14\omega_x^2$
are computed from the two-band model.
The frequency scale $\omega_F= F(0)/\beta \alpha$
is fixed by the sum-rule (\ref{sumrule}).

\begin{figure}
\epsfxsize7.5cm
\epsfbox{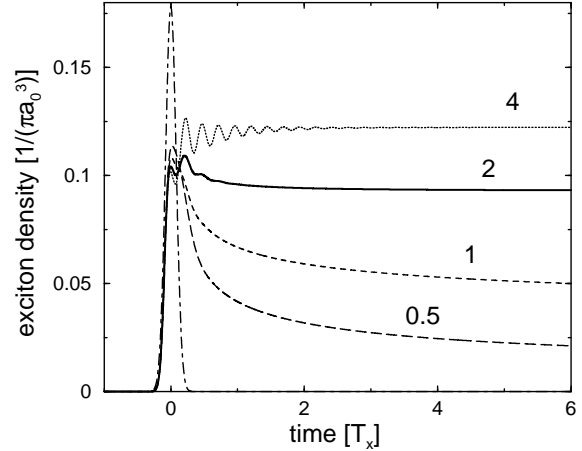}
  \caption{\label{f3}
    Coherent polarization for the three-dimensional semiconductor model with
    $\omega_x=$ 4.2~meV, pulse width 200~fs and pulse area $4\pi$
    for various $\alpha$ of the model correlation function.
    The other parameters are given in the text.
    The dashed-dotted line is the excitation pulse.}
\end{figure}

Figure~\ref{f3} shows the numerical solution of Eq.~(\ref{main}) for the model
correlation with various values for the parameter $\alpha$ for an excitation
with a 200~fs laser pulse and area $4\pi$. For $\alpha\ge 2$, the asymptotic
behavior of the exciton density changes to constant from  decaying.
For  $\alpha\ge 1$, small oscillations are superimposed on the slow density
evolution. The period of  oscillations is about 200~fs for $\alpha=4$.
These oscillations do not correspond to the frequency peak
${\omega}_c=F(0)/\beta$ of the spectral density, which does not vary with
increasing $\alpha$ as the peak narrows.
They are collective modes in density fluctuation excited by the strong
$4\pi$-pulse. They do not occur in the slowly varying approximation
(\ref{markoff}). It is, however, possible to treat the modes as a small,
fast-varying perturbation $P_1$ to the slow component $P_0$.  Inserting the
combination $P=P_0 + P_1$ in Eq.~(\ref{main}) and keeping $P_1$ only to first
order, we find
\begin{eqnarray}  \label{linear}
  \dot{P_1} &=&  - |P_0|^2 \left[ i \beta  P_1
   +\int_{0}^t F(t-t') P_1(t') dt' \right] \\
&-& \frac{P_0^2}{2} g(t)  \nonumber
 P_1^\ast(t).
\end{eqnarray}
The last contribution of Eq.~(\ref{linear}), being
$\dot{P_0} P_1^\ast/P_0^\ast$ is negligible compared with $\dot{P_1}$.
The resulting linear equation can be solved by Fourier transform
of $ P_1(t)$ for a slowly-varying density to
yield  the collective-mode frequencies as zeros of the equation
\begin{eqnarray}  \label{mode}
\omega &=& n \Sigma(\omega) , \\
  \Sigma(\omega) &=&  \beta - F(0) \int_0^{\infty}
    \frac{\rho(\omega') d\omega'}{\omega' - \omega -i0} .
\nonumber
\end{eqnarray}

\begin{figure}
\epsfxsize7.5cm
\epsfbox{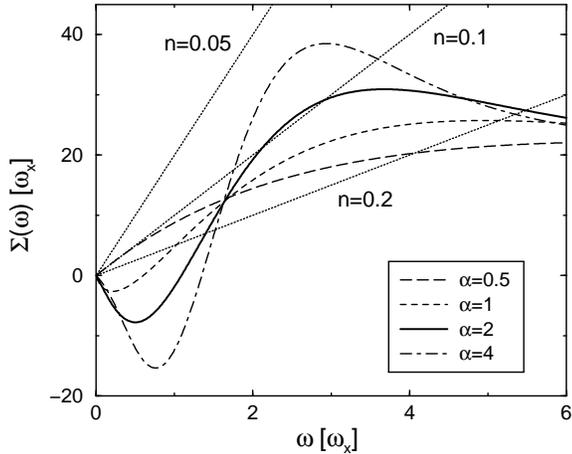}
\caption{\label{f4}
  Graphical solution of the collective mode equation
  (\protect{\ref{mode}}) for three different exciton densities
  of $n=$ 0.05, 0.1 and 0.2 for the parameters of Fig.~\protect{\ref{f3}}.
  The fine-dotted line corresponds to $\omega/n$.}
\end{figure}

 Figure~\ref{f4} shows a graphical solution
of the collective mode equation from the crossings between the real part of
$\Sigma$ and the straight line $\omega/n$.  The $\omega=0$ solution
corresponds to the Goldstone mode\cite{KK}. Models for a range of $\alpha$
parameters are shown.   Above a critical density, e.g., $n\approx 0.1$ (in the
density units of Fig.~\ref{f3}) for
$\alpha= 2$, additional solutions with two high frequencies exist,
but only the higher frequency solution with the weaker damping corresponds to
the oscillating mode in Fig.~\ref{f3}.

The imaginary part of $\Sigma$ represents the damping of the collective mode by
the two-exciton continuum.  It peaks around $\omega=2\omega_x$ where the real
part rises steeply and, thus, damps out the lower frequency solution.
 We have not included the damping by two
excitons when one (or both) of them is dissociated into an
electron-hole pair in our computation but
this would not qualitatively change the damping continuum.
In the three-dimensional model where $T_x\approx 1$~ps,
these oscillations have periods between 50--200~fs.  The curvature in
the $\alpha=0.5$ case is exceptional, resulting in only one high-frequency
solution whose damping goes down with increasing excitation density.

In conclusion, we extended the Gross-Pitaevskii equation for the exciton
polarization at low densities
 to include correlation, which is an important
counterpart to the mean-field exchange term and which provides an intrinsic
dephasing mechanism.  A consequence is the existence of collective oscillations
riding on a slowly varying density.  The Goldstone mode, though undetectable,
indicates a Bose-Einstein condensate behavior, induced by optical excitation
but sustained by exciton interaction.

Finding by nonlinear optical measurement of exciton polarization
of the high-frequency mode, predicted by the same dispersion
relation as the Goldstone mode, could then be argued to present evidence of
condensation.

We thank D. Chemla, M. Cohen, and S. Louie  for the
hospitality at Lawrence Berkeley National Laboratory and the University of
California, Berkeley. LJS thanks  the Miller Institute for a visiting
professorship.  This work was supported in part by the Director, Office of
Science, Office of Basic Energy Sciences, Materials Sciences Division of
the U.S.
Department of Energy under Contract No. DE\#-AC03-76SF00098, in part by the
Sonderforschungsbereich SFB
$345$, G\"ottingen, and in part by the NSF Grant No. DMR 9721444. We thank D.
Chemla, S. Bolton, and U. Neukirch for helpful discussions.
%

%-------------------------- captions ---------------------------------


\begin{references}
\vspace{-1.cm}
\bibitem[\ast]{goe} Permanent address: Institut f\"ur Theoretische Physik,
Universit\"at G\"ottingen, Bunsenstra{\ss}e 9, D-37073 G\"ottingen,
Federal Republic of Germany.

\bibitem[\dagger]{lajolla} Permanent address:
Department of Physics, University of California
San Diego, La Jolla, California 92093-0319.

%--------------------------------------------------------------

\bibitem{rabi} I. I. Rabi, Phys. Rev. {\bf 51}, 652 (1937).

\bibitem{hahn} S. McCall and E.L. Hahn, Phys. Rev. Lett. {\bf 18},
908 (1967); Phys. Rev. {\bf 183}, 457 (1969).

\bibitem{allen} L. Allen and J. H. Eberly,
{\it Optical Resonance and Two-Level
Atoms} (Wiley and Sons, New York, 1975).

\bibitem{shah} J. Shah, {\em Ultrafast spectroscopy of semiconductors and
semiconductor nanostructures}, (Springer, Berlin, 1996).

\bibitem{chemla} D.S. Chemla, Semiconductors and Semimetals {\bf 38}, 175
(1999).

\bibitem{binder} R. Binder {\em et al.},
%S.W. Koch, M. Lindberg, N. Peyghambarian,
%and W. Sch\"afer,
Phys. Rev. Lett. {\bf 65}, 899 (1990).

\bibitem{exact} A. Knorr {\em et al.},
%Th. \"Ostreich, K. Sch\"onhammer, R. Binder,
%and S.W. Koch,
Phys. Rev. B {\bf 49}, 14024 (1994).

\bibitem{schul} A. Sch\"ulzgen {\em et al.},
%R. Binder, M. E. Donovan, M. Lindberg,
%K. Wundke, H. M. Gibbs, G. Kitrova, and N. Peyghambarian,
Phys. Rev. Lett. {\bf 82}, 2346 (1999).

\bibitem{gross} E.P. Gross, Nuovo Cimento, {\bf 20}, 454 (1961).

\bibitem{pita} L.P. Pitaevskii, Sov. Phys. JETP {\bf 13}, 451 (1961).

\bibitem{hana} E. Hanamura, Solid State Commun. {\bf 91}, 889 (1994).

\bibitem{talanina} I. Talanina {\em et al.},
%D. Burak, R. Binder, H. Giessen,
%and N. Peyghambarian,
Phys. Rev. E {\bf 58}, 1074 (1998).

\bibitem{giess} H. Giessen {\em et al.},
%A. Knorr, S. Haas, S. W. Koch, S. Linden,
%J. Kuhl, M. Hetterich, M. Gr\"un, and C. Klingshirn,
Phys. Rev. Lett. {\bf 81}, 4260 (1998).

\bibitem{KK} L.V. Keldysh and A.N. Kozlov,
Sov. Phys. JETP {\bf 27}, 521 (1968).

\bibitem{review} Th. \"Ostreich, K. Sch\"onhammer, and L. J. Sham,
Phys. Rev. Lett. {\bf 74}, 4698 (1995);
Phys. Rev. B{\bf 58}, 12920 (1998).

\bibitem{avk}  An alternative derivation of the same memory function as in
Ref.~\onlinecite{review} has been given by V.M. Axt, K. Victor, and T. Kuhn,
Phys. Stat. Sol. (b) {\bf 206}, 189 (1998).

\bibitem{eid} H. Wang {\em et al.},
%K.B. Ferrio, D.G. Steel, P.R. Berman, Y.Z. Hu,
%R. Binder, and S.W. Koch,
Phys. Rev. A {\bf 49}, 1551 (1994).

\bibitem{axt} V.M. Axt and A. Stahl, Zeitschrift f\"ur Physik B {\bf 93},
195 (1994); 205 (1994).


\bibitem{diag1}  M. Z. Maialle and L. J. Sham, Phys. Rev. Lett., {\bf
    73}, 3310 (1994).


\bibitem{exho} M. Wegener, D. S. Chemla, S. Schmitt-Rink, and W.
  Sch\"afer, Phys. Rev. A {\bf 42}, 5675 (1990).

\bibitem{non} Th. \"Ostreich and A. Knorr, Phys. Rev. B {\bf 50}, 5717
(1994).

\bibitem{kner} P. Kner {\em et al.},
%S. Bar-Ad, M.V. Marquezini, D.S. Chemla,
%and W. Sch\"afer,
Phys. Rev. Lett., {\bf 78}, 1319 (1997).

\bibitem{rules} Th. \"Ostreich and L. J. Sham, to be published.

\end{references}
\end{document}